\documentclass[aps,prl,twocolumn,groupedaddress,floatfix,superscriptaddress]{revtex4-1}

\usepackage{amsmath}
\usepackage{amssymb}
\usepackage{graphicx}
\usepackage{float}
\usepackage[english]{babel}
\usepackage{dsfont}
\usepackage{epstopdf}
\usepackage{soul}
\usepackage{color}
\usepackage{braket}
\usepackage[colorlinks=true,
            linkcolor=magenta,
            urlcolor=blue,
            citecolor=blue]{hyperref}

\usepackage[normalem]{ulem}

\def\be{\begin{equation}}
\def\ee{\end{equation}} 
\def\bea{\begin{eqnarray}}
\def\eea{\end{eqnarray}} 
\def\ba{\begin{array}} 
\def\ea{\end{array}}
\def\pa{\partial}

\def\om{\omega}

\def\nn{\nonumber}
\def\ket{\rangle}
\def\bra{\langle}
\def\b{\mathbf}
\def\bs{\boldsymbol}
\def\f{\frac}

\def\ra{\rightarrow}
\def\rm{\mathrm}

\newcommand{\mv}[1]{\langle #1\rangle}

\definecolor{Nathanblue}{rgb}{0.92,0.24,0.24}

\emergencystretch=\maxdimen
\begin{document}
\title{Chiral orbital order of interacting bosons without higher bands}
\author{Marco Di Liberto}
\email{marco.diliberto@unipd.it}
\affiliation{Dipartimento di Fisica e Astronomia ``G. Galilei" \& Padua Quantum Technologies
Research Center, Università degli Studi di Padova, I-35131, Padova, Italy}
\affiliation{INFN Istituto Nazionale di Fisica Nucleare, Sezione di Padova, I-35131, Padova, Italy}
\affiliation{Institute for Quantum Optics and Quantum Information of the Austrian Academy of Sciences, Innsbruck, Austria}
\author{Nathan Goldman}
\email{nathan.goldman@ulb.be}
\affiliation{Center for Nonlinear Phenomena and Complex Systems, Universit\'e Libre de Bruxelles, CP 231, Campus Plaine, B-1050 Brussels, Belgium}

\begin{abstract}

Ultracold atoms loaded into higher Bloch bands provide an elegant setting for realizing many-body quantum states that spontaneously break time-reversal symmetry through the formation of chiral orbital order. 
The applicability of this strategy remains nonetheless limited due to the finite lifetime of atoms in high-energy bands. 
Here we introduce an alternative framework,  suitable for bosonic gases, which builds on assembling square plaquettes pierced by a $\pi$-flux (half a magnetic-flux quantum). 
This setting is shown to be formally equivalent to an interacting bosonic gas loaded into $p$ orbitals, and we explore the consequences of the resulting chiral orbital order, both for weak and strong onsite interactions. 
We demonstrate the emergence of a chiral superfluid vortex lattice, exhibiting a long-lived gapped collective mode that is characterized by local chiral currents. 
This chiral superfluid phase is shown to undergo a phase transition to a chiral Mott insulator for sufficiently strong interactions. 
Our work establishes coupled $\pi$-flux plaquettes as a practical route for the emergence of orbital order and chiral phases of matter.

\end{abstract}

\maketitle

%\section{Introduction}

\emph{Introduction.} Breaking time-reversal symmetry is known to drastically alter the phases and dynamical properties of quantum matter, as was evidenced by vortex lattices in rotating ultracold gases~\cite{madison2000vortex,abo2001observation,Cooper_review} and the quantum Hall effects in 2D materials immersed in strong magnetic fields~\cite{girvin2005introduction}. In the context of cold atoms in optical lattices, this fundamental symmetry can be broken by privileging a certain orientation of motion, e.g.~by rotating the system~\cite{Tung_rotating_lattice,Cooper_review} or by applying a circular shaking to the lattice~\cite{struck2013engineering,jotzu2014experimental}. However, these methods lead to instabilities or heating, hence complicating the formation of strongly-correlated phases~\cite{Cooper_review,RevModPhys.91.015005}. This motivates the development of alternative schemes to break time-reversal symmetry in ultracold gases. 

A first possible route builds on addressing different internal states of an atom with lasers, in view of engineering synthetic lattice structures with effective magnetic fluxes~\cite{RevModPhys.83.1523,goldman2014light,jaksch2003creation,gerbier2010gauge,PhysRevLett.106.175301,PhysRevLett.112.043001,Browaeys2020}.
Such schemes have been experimentally implemented and lead to the observation of chiral states and dynamics \cite{atala2014observation,stuhl2015visualizing,mancini2015observation,an2017direct,chalopin2020probing}.
A second route exploits the orbital structure of higher Bloch bands \cite{Girvin2005, Liu2006} in the absence of external driving. 
In these systems, interactions couple the different orbitals and can favour ground states with finite angular momentum, thus spontaneously breaking time-reversal symmetry~\cite{Li2016}.
The realization of higher-band lattice models offers the opportunity to observe intriguing correlated quantum phases with chiral properties \cite{Wu2006, Wu2008, Wu2008b, Zhao2008, Sun2012, Li2012, Pinheiro2013, Li2021} as well as topological chiral excitations \cite{DiLiberto2016, Xu2016}.
Quantum gases have been experimentally loaded into $p$ bands \cite{Bloch2007,Wirth2011}, as well as higher bands \cite{Olschlager2011}, hence leading to the observation of short-lived condensates with spontaneously broken time-reversal symmetry~\cite{Wirth2011,Xu2021,Li2021b}. 
However, atoms in higher bands have a relatively short lifetime due to atom-atom collisions, thus making it challenging to reach the strongly-interacting regime.

%%%%%%%%%  Figure  %%%%%%%%%%%%%%%%%%%%%%%%%
\begin{figure}[!t]
\center
\includegraphics[width=0.95\columnwidth]{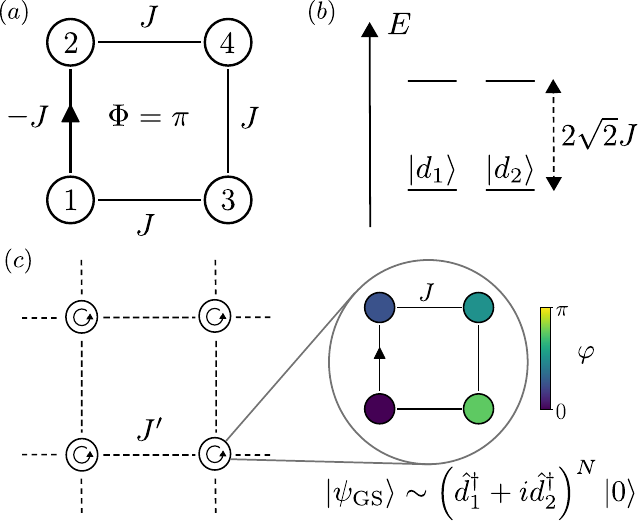}

\caption{The $\pi$-flux plaquette building block for chiral bosonic phases. (a) Schematic representation of a single plaquette with flux $\Phi=\pi$. 
(b) Single-particle spectrum displaying two-fold degeneracy separated by an energy gap. (c) Sketch of an extended lattice of connected building-blocks, with $J' \ll J$, resulting in a chiral superfluid phase:~each super-site hosts a quantized vortex associated with the condensate phase winding, $\varphi_i = \text{arg}\mv{\hat b_i} $. 
}
\label{fig:intro}
\end{figure}
%%%%%%%%%%%%% End Figure %%%%%%%%%%%%%%%%%%%%%

In this work, we introduce an alternative route to $p$-band orbital physics, which does not rely on populating higher bands.
Our approach is based on an analogy between $p_{x,y}$ orbitals and the two-fold degenerate low-energy orbitals $d_{1,2}$ provided by a square plaquette pierced by a $\pi$-flux [Figs.~\ref{fig:intro}(a)-(b)].
Considering bosons with Hubbard interactions, we show that this setting favors chiral orbital order and breaks time-reversal symmetry.
Our construction consists in coupling these $\pi$-flux plaquettes with weak links, such as to preserve the orbital order over extended lattices [Fig.~\ref{fig:intro}(c)]. We explore this approach both in the weakly and strongly-interacting regimes.
In particular, we demonstrate the emergence of a chiral superfluid vortex phase, exhibiting a long-lived gapped collective mode and local chiral currents. 
Furthermore, we show that the superfluid undergoes a transition to a chiral Mott insulator for sufficiently strong interactions and half filling. 
Our results establish coupled $\pi$-flux plaquettes as powerful building blocks for orbital-like physics and chiral phases of matter.

%\section{Single-plaquette with $\pi$-flux: Model and effective theory}

\emph{Single-plaquette with $\pi$-flux:~Effective theory.} 
The interplay of magnetic flux and interactions can have remarkable effects, even at the level of a single plaquette \cite{Roushan2017, Browaeys2020, Pollman2020}.
Here, we start by considering a single square plaquette pierced by a $\pi$ flux [see sketch in Fig.~\ref{fig:intro}(a)], as described by 
\be
\hat H_0 = -J \left(e^{i\pi}\hat b_1^\dag \hat b_2^{} +  \hat b_2^\dag \hat b_4^{} + \hat b_4^\dag \hat b_3^{} + \hat b_3^\dag \hat b_1^{} + \mathrm{H.c.}\right) \,,
\ee
with onsite Hubbard interactions among the bosons
\be 
\hat H_{\mathrm{int}} = \f{U}{2} \sum_i \hat n_i (\hat n_i - 1)\,,
\ee 
and we set $U>0$ (repulsive).
The full Hamiltonian reads $\hat H = \hat H_0 + \hat H_{\mathrm{int}}$.
The Hamiltonian $\hat H_0$ admits four eigenstates that are pairwise degenerate in energy, $\epsilon_{1,2} = -\sqrt 2 J$ and $\epsilon_{3,4} = \sqrt 2 J$; see Fig.~\ref{fig:intro}(b). 
We indicate the operators corresponding to the modes with eigenvalues $\epsilon_i$ as $\hat d_i^{}$, $\hat d_i^{\dag}$, defined through the unitary transformation $\hat b_i^{} = \sum_{ij} \mathcal{U}_{ij} \hat d_j^{}$~\cite{DiLiberto2020,SuppMat}. 
Since the single-particle theory displays an energy gap $\Delta \epsilon =  2\sqrt 2 J$, it is meaningful to consider the projected Hamiltonian $\hat H_{\text{eff}} \equiv \hat P \hat H \hat P$, where $\hat P$ is the projection operator onto the low-energy subspace spanned by the modes with $i=1,2$ \cite{Huber2010,Cooper2012}:
\be
\label{eq:HamP}
\hat H_{\text{eff}} = \f{3U}{16} \hat n^2  - \f{U}{16} \hat L_z^2 - \left(\sqrt 2 J + \f U 8 \right) \hat n\,,
\ee
where $\hat n = \hat d_1^\dag \hat d_1^{} + \hat d_2^\dag \hat d_2^{}$ and $\hat L_z = i (\hat d_1^\dag \hat d_2^{} - \hat d_2^\dag \hat d_1^{})$. 
This expression shows that the low-energy physics of weakly-interacting bosons in a $\pi$-flux plaquette shares similarities with $p$-band bosons \cite{Liu2006}:~In direct analogy with $p_{x,y}$ orbitals, the two modes $d_1$ and $d_2$ experience density-density repulsive interactions but also an orbital-like coupling $-\!\hat L_z^2$ with a \emph{negative} sign.
The operator $\hat L_z$ has the same structure as the angular momentum operator built from $p_{x,y}$ orbitals, and the negative sign  privileges a ground state with the highest angular momentum possible (for $U>0$), i.e.~a macroscopic occupation of a complex orbital \mbox{$\vert d_1\rangle \pm i \vert d_2 \rangle$} \cite{Liu2006}.
Besides the global $U(1)$ symmetry associated with the conservation of the total number of particles and time-reversal symmetry, the low-energy Hamiltonian displays an emergent discrete $\mathbb Z_2$ symmetry represented by $\hat d_1^{} \!\ra\! \hat d_2^{}$ and $\hat d_2^{} \!\ra\! \hat d_1^{}$, which transforms the angular momentum as $\hat L_z \ra -\hat L_z$.

The Hamiltonian in Eq.~\eqref{eq:HamP} can be solved exactly by noting that $[\hat L_z, \hat H_\text{eff}]=0$.
Indicating the single-particle eigenstates of $\hat L_z$ as $|\pm \ket \equiv \hat d^\dag_\pm |0\ket$, with  $\hat L_z |\pm\ket = \pm |\pm\ket$ and $\hat d^\dag_\pm = (\hat d^\dag_1 \pm i \hat d^\dag_2)/\sqrt 2$, a generic many-body eigenstate  with energy $E_{\rm{eff}}$ can therefore be written as 
\begin{align}
\label{eq:exact_eigs}
&|n_+, n_- \ket = \f{1}{\sqrt{n_+!\, n_-!}}(\hat d_+^\dag)^{n_+} (\hat d_-^\dag)^{n_-} |0\ket\,,\\
&E_{\rm{eff}}(n_+,n_-) = \frac{3UN^2}{16} \!-\! \frac{U}{16}(n_+ \!-\! n_-)^2 \!-\! \left(\sqrt 2 J \!+\! \frac{U}{8} \right) N\, ,\notag 
\end{align}
with $N \!=\! n_+ + n_-$. 
The two-fold degenerate ground state thus corresponds to $n_+\!=\!N$ or $n_-\!=\!N$, i.e.~$|\psi_{\text{GS}}\rangle \sim \left(\hat d_1^\dag \pm i \hat d_2^\dag \right)^N |0\rangle$, with energy per particle $E_{\rm{GS}}/N = g/8 - \sqrt 2 J - g/8N$.
Such a ground state, which breaks the aforementioned time-reversal and $\mathbb Z_2$ symmetries~\footnote{Notice that the two symmetries (time-reversal and orbital $\mathbb Z_2$) are not equivalent. 
If the Hamiltonian had a term $\delta E\, \hat d_1^\dag \hat d_1^{}$, this would explicitly break the $\mathbb Z_2$ orbital symmetry while preserving time-reversal. 
However, for sufficiently small $\delta E$, the ground-state would still be a time-reversal broken state $\left(\sin \alpha\, \hat d_1^\dag \pm i \cos \alpha\, \hat d_2^\dag \right)^N |0\rangle$, where $\alpha$ can be expressed in terms of $\delta E$.},  is a chiral condensate with angular momentum $\mv{\hat L_z} = \pm N$.
Let us consider the ground state with positive chirality, $n_+\!=\!N$. 
The lowest one-particle excitation, which is given by $n_+ = N - 1$ and $n_- = 1$, corresponds to removing a particle from the condensate and transferring it to a state with opposite angular momentum.
The energy corresponding to this elementary excitation is $\Delta E_{\text{exc}} =  g / 4 - U / 4$.

%\section{Mean-field solution}

\emph{The collective mode on a single plaquette.} To gain more insight on this excitation, we now discuss the weakly-interacting regime within a hydrodynamic approach relevant to systems of ultracold atoms and nonlinear photonics. 
We therefore consider a finite $g\equiv U N \ll J$, and take the limit \mbox{$N\ra \infty$}. 
Under these assumptions, the problem is treated using a discrete Gross-Pitaevskii description with $N$ condensed particles: we replace the operators in Eq.~\eqref{eq:HamP} by the ansatz
\be 
\hat d_1 \ra \mv{\hat d_1} = \sqrt{N_1} \,, \quad \hat d_2 \ra \mv{\hat d_2} = e^{i\theta}  \sqrt{N_2}\,,
\ee 
with $N \!=\! N_1 \!+\! N_2$, and construct the energy functional
\be
 E_{\text{MF}}[d_1,d_2] = E(N) -\f U 4 N_1 (N-N_1) \sin^2\theta \,,\label{EMF_eq}
\ee
where $E(N) = 3UN^2/16  - \left(\sqrt 2 J + U/8 \right) N$. Note that the last term in Eq.~\eqref{EMF_eq},  which corresponds to $-\hat L_z^2$, is minimized for $N_{1,2} \!=\! N/2$ and $\theta \!=\! \pm \pi/2$, thus breaking time-reversal symmetry. This yields the ground state energy per particle $E_{\textrm{GS}}^{\textrm{MF}}/N\!=\! -\sqrt 2 J \!+\! g/8$, which is in agreement with $E_\text{GS}/N$ for $N\ra\infty$.

The eigenmodes of the condensate can be obtained by studying the fluctuations with respect to the stationary solution. 
We introduce the Lagrangian density
\be
\label{eq:lagrangian}
\mathcal{L} = i (d_1^* \pa_t d_1^{} + d_2^* \pa_t d_2^{}) - E_{\text{MF}}[d_1,d_2]\,,
\ee
and define small fluctuations within a hydrodynamic picture as $N_{1,2} \ra N/2 \pm \delta \rho$ and $\theta \ra \pi/2 +
\delta \theta$. At lowest order, we obtain $E_{\text{MF}}[d_1,d_2] = E^{(0)}_{\textrm{MF}} + \delta E^{(2)}_{\textrm{MF}} $, where
\be
\delta E_{\text{MF}}^{(2)} = \f{UN^2}{16} \delta\theta^2 + \f U 4 \delta \rho^2\,.
\ee
The dynamical variables are the relative density and the relative phase, satisfying the equations of motion $\pa_t \delta\theta = \f{U}{2} \delta\rho$ and $\pa_t \delta\rho = -\f{UN^2}{8} \delta\theta$,
which have the solutions
\be
\delta\theta = A \cos \om_0 t \,, \quad \delta\rho = -\f{A N}{2} \sin \om_0 t\,,\label{eq_solutions_motion}
\ee
with $\omega_0 \!=\! g/4$ and $A$ an arbitrary constant set by the initial conditions. As a distinctive signature of the mode, the oscillation of the two conjugate variables occurs with a phase difference $\pi/2$. 
The nature of this mode shares similarities with a recent measurement in $p$ bands \cite{Paramekanti2014,VargasHemmerich2021}.

%\section{External driving}

%%%%%%%%%  Figure  %%%%%%%%%%%%%%%%%%%%%%%%%
\begin{figure}[!t]
\center
\includegraphics[width=0.95\columnwidth]{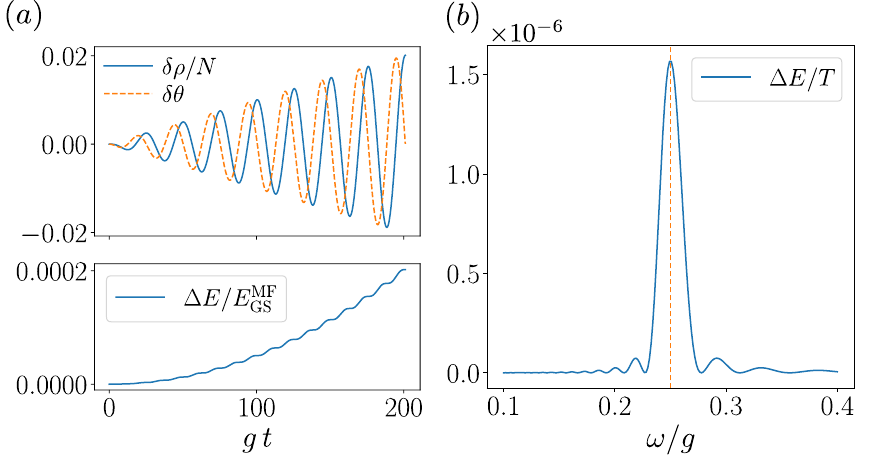}

\caption{Exciting the collective mode by driving.  (a) Phase and density dynamics at resonance $\om = \omega_0 = g/4$ and corresponding energy absorption for $V_0 = 10^{-4}g$. (b) Resonance peak obtained by measuring the absorption energy rate over $t=10T$, where $T=2\pi/\om$. The vertical dashed line is drawn at the expected resonance condition $\omega = g/4$.
}
\label{fig:driving}
\end{figure}
%%%%%%%%%%%%% End Figure %%%%%%%%%%%%%%%%%%%%%

\emph{Detecting the collective mode.} We now show how to detect the collective mode via spectroscopy or quench protocols. 
We first consider a time-periodic modulation of the onsite energy at two opposite corners of the plaquette, 
\begin{align}
\label{eq:modulation}
\delta \hat V &= V(t) (\hat b_2^\dag \hat b_2^{} - \hat b_3^\dag \hat b_3) \nn \\
&\approx \f{V(t)}{2} (\hat d_2^\dag \hat d_2^{} - \hat d_1^\dag \hat d_1) \ra - V(t) \delta \rho\,,
\end{align}
with $V(t) = V_0 \sin \omega t$. 
As shown by Eq.~\eqref{eq:modulation}, this spectroscopic probe couples to the relative density $\delta \rho$ within the projected theory.
When reaching resonance, $\omega \!=\! \omega_0 \!=\! g/4$, the system is expected to absorb energy while populating the collective mode characterized by Eq.~\eqref{eq_solutions_motion}. 
This is confirmed in Fig.~\ref{fig:driving}(a), which shows a numerical integration of the full nonlinear equations of motion, obtained from the Lagrangian \eqref{eq:lagrangian} with the addition of the drive $\delta\hat V(t)$. 
One recognizes the characteristic $\pi/2$ phase difference between the relative density $\delta \rho$ and relative phase $\delta \theta$ [Eq.~\eqref{eq_solutions_motion}]. 
In Fig.~\ref{fig:driving}(b), the energy absorption per unit period of driving shows a peak at $\omega = \omega_0$, confirming our analysis.

%%%%%%%%%  Figure  %%%%%%%%%%%%%%%%%%%%%%%%%
\begin{figure}[!b]
\center
\includegraphics[width=0.99\columnwidth]{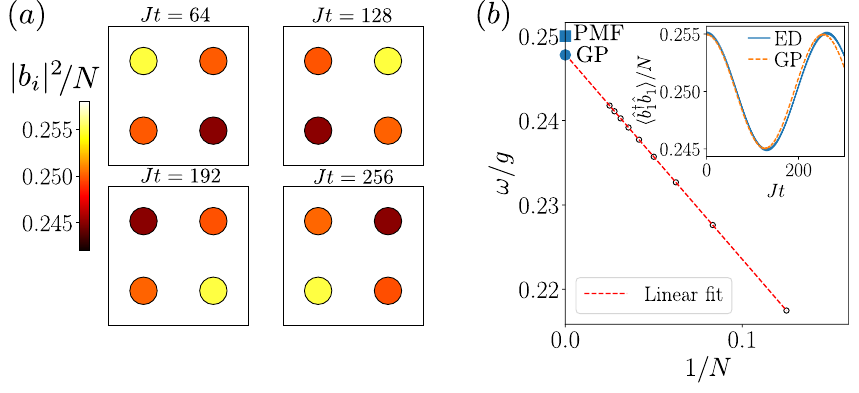}

\caption{Chiral impurity dynamics.  (a) Snapshots of the density distribution  at different times showing the chiral motion of the impurity for $\Delta = 0.001 J$ and $g=0.1 J$ obtained from the full GP dynamics.
(b) Oscillation frequency for $g=0.1J$ obtained by ED for different particle numbers $N$ and scaling for $N\rightarrow\infty$. The solid square indicates the projected mean-field (PMF) theory result $g/4$ and the solid circle indicates the GP oscillation frequency. In the inset, single-site density dynamics obtained within the GP description compared with ED for $N=32$ and $U = g/N$.
}
\label{fig:impurity}
\end{figure}
%%%%%%%%%%%%% End Figure %%%%%%%%%%%%%%%%%%%%%

%\section{Impurity dynamics}

The exact solution in Eq.~\eqref{eq:exact_eigs} already revealed that the excited mode corresponds to a single-particle excitation carrying angular momentum. 
This translates into a real-space current that can be detected by the following quench protocol.
Inspired by the dynamics of density defects in topological systems  \cite{Spielman2013, Goldman2013, Pollmann2018}, we introduce a small onsite ``impurity" potential $\Delta \hat H = - \Delta\, \hat b_1^\dag \hat b_1^{}$, with $\Delta>0$ and $\Delta \ll g \ll J$.
This creates an initial state with a small excess density on one site. 
The energy functional $E_{\text{MF}}(\theta)/\rho = (g/32) \cos 2 \theta + (\Delta/4)\cos \theta$ shows that this state corresponds to occupying a small fraction of the excited mode (of the unperturbed system) with $\theta = \arccos(-2\Delta/g)\neq\pm\pi/2$. 
At $t\!=\!0$, we then quench $\Delta \ra 0$ and let the system evolve in time.
The real-space dynamics is obtained by directly solving the 4-sites Gross-Pitaevskii (GP) equations $i\pa_t b_i = -\sum_j J_{ij} b_j + U |b_i|^2 b_i$; it displays a clear chiral motion of the impurity, as shown in Fig.~\ref{fig:impurity}(a).
These results were benchmarked by computing the dynamics of the full many-body system, using exact-diagonalization (ED)~\cite{QuSpin2019}. Figure~\ref{fig:impurity}(b) shows a scaling analysis of the extracted oscillation frequency in the $N\ra\infty$ limit, confirming the GP results at short times.
We attribute the small mismatch with the projected theory result ($\omega = g/4$) to the perturbative contribution of the high-energy orbitals. 

%\section{Extension to 2D: the BBH model}

%%%%%%%%%  Figure  %%%%%%%%%%%%%%%%%%%%%%%%%
\begin{figure}[!b]
\center
\includegraphics[width=\columnwidth]{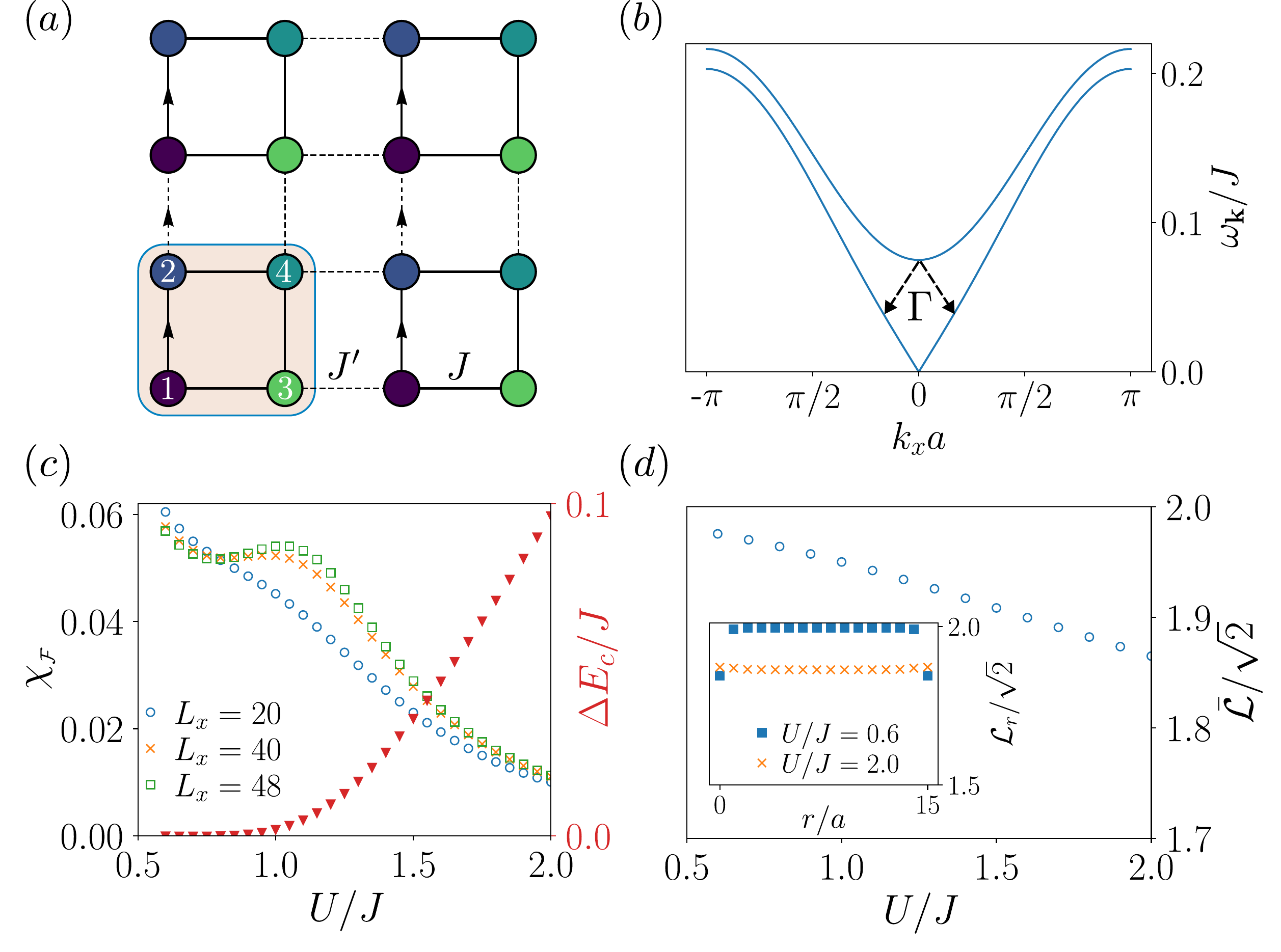}

\caption{Interacting BBH model. (a) Tight-binding representation of the BBH model with uniform $\pi$-flux and staggered hopping amplitudes $J$, $J'$. The color of each site represents the ground state phase pattern for the weakly-interacting superfluid vortex lattice (units as in Fig.~\ref{fig:intro}). 
(b) Bogoliubov spectrum within the projected theory for $J' = 0.1J$ and $g=0.1J$, showing the appearence of a gapless and a gapped mode at small momenta. Dashed lines indicate the decay channel with rate $\Gamma$ for the vortex excitation discussed in the main text. 
(c) Fidelity susceptibility $\chi_{_\mathcal F}$ for a ladder with $L_y=2$ sites and up to $L_x = 48$ sites, extracted with DMRG for $J'=0.1J$. The peak indicates a phase transition, from a superfluid to an insulating phase, as further evidenced by the finite charge gap $\Delta E_c$ obtained by finite-size scaling. 
(d) Averaged loop-current $\bar{\mathcal L}\equiv \sum_{r}|\mv{\hat{\mathcal L}_{r}}|/N_p$ over the ladder plaquettes for $L_x = 32$, showing that time-reversal symmetry remains broken across the transition. The inset shows two profiles of loop-current patterns across the transition.
}
\label{fig:BBH}
\end{figure}
%%%%%%%%%%%%% End Figure %%%%%%%%%%%%%%%%%%%%%

\emph{Chiral phases in extended lattices.}
Having established the chiral properties emerging from bosonic $\pi$-flux plaquettes, we now demonstrate how assembling these building blocks can lead to chiral phases of matter, both in the weakly and strongly interacting regimes. 
Our construction consists in connecting $\pi$-flux plaquettes with weak couplings $J'\ll J$, such that each plaquette can be viewed as a super-site hosting two orbitals (in direct analogy with $p$-band models \cite{Hu2015, Amo2015}). 
Considering a square geometry [Fig.~\ref{fig:BBH}(a)], this naturally leads us to an interacting bosonic version of the BBH model \cite{Benalcazar2017}, initially introduced in the context of higher-order topological insulators.

We define the projected orbital operators acting on each plaquette as $\hat d_{1,\b r}$ and $\hat d_{2,\b r}$, where $\mathbf r$ indicates the plaquettes position. 
The local terms describing the dynamics within each plaquette are represented by the Hamiltonian \eqref{eq:HamP}. 
Upon projection ($\hat P$), the new inter-plaquette terms ($J'$) are described by
\be
\label{eq:projJ1}
\hat H_{\text{eff}}' = - \f{J'}{2 \sqrt 2} \sum_{\b r, \sigma, \nu}  \left( \hat d^{\dag}_{\sigma, \b r} \hat d^{}_{\sigma,\b r + \b e_{\nu}} + \textrm{H.c.} \right) \,,
\ee
where $\b e_{x} \!=\! (a, 0)$, $\b e_{y} \!=\! (0,a)$ and $a$ is the lattice constant. 
In the weakly-interacting limit, the ground state is a uniform condensate ($\bs\Gamma$ point) forming a superfluid vortex lattice \cite{LKLim2008}, with the condensate phase pattern shown in Fig.~\ref{fig:BBH}(a); this simply repeats the single plaquette result indicated in Fig.~\ref{fig:intro}.
To study the fate of the chiral mode identified in a single plaquette, we perform a Bogoliubov analysis. 
We obtain the Bogoliubov Hamiltonian $\hat H^{\textrm{Bog}} = \sum_{\b k\neq 0, \alpha} \om_{\alpha,\b k}\,  \hat \beta^{\dag}_{\alpha,\b k} \hat \beta^{}_{\alpha,\b k}$, with $\alpha=1,2$. 
Here $\hat \beta^{\dag}_{\alpha,\b k}$ creates a Bogoliubov quasiparticle and the spectrum reads  $\om_{1,\b k} = \sqrt{\xi_{\b k}(\xi_{\b k}+ g )}$ and $\om_{2,\b k} = g/4 +  \xi_{\b k}$,
with $\xi_{\b k} \equiv -J'(\cos k_xa + \cos k_ya - 2)/\sqrt 2 $. 
The two branches of the spectrum are shown in Fig.~\ref{fig:BBH}(b).
The massless (Goldstone) mode at small momenta becomes $\om_{1,\b k} \approx c_s |\b k| $ with the sound velocity $c_s = \sqrt{J' g/4\sqrt{2}}$, while the massive mode has a gap $g/4$, which corresponds to the single-plaquette gapped mode of energy $\om_0$. 

In the limit of disconnected plaquettes, $J'\ll g$, the excited mode is long lived as there is no lower energy mode to decay to. 
As the coupling strength approaches \mbox{$J' \sim g$}, the massive mode becomes energetically resonant with the gapless phonon branch, thus opening a direct channel for its decay.
Within the effective projected theory, we computed the couplings $\hat H^{(3)}$ between the massive and massless modes up to cubic powers of $\delta \hat d_{\alpha,\b k}^{(\dag)}$. 
Assuming $g\ll J'$, the decay rate of the massive mode can be estimated as $\Gamma = \pi \sum_{\b q} \delta (\om_0 - 2 \epsilon_{1,\b q})\, |\bra \beta_{1,\b q} \beta_{1,-\b q} | \hat H^{(3)} |  \beta_{2, \b 0} \ket|^2 = 0.73 \pi^2 g U / (512 J')$~\cite{SuppMat}, thus suggesting that the mode is well defined  ($\Gamma \lesssim \om_0 = g/4$) under the condition $U \lesssim 17.85 J'$.

From the effective (projected) model, and for appropriate filling factors, we expect a phase transition from the chiral superfluid to a chiral Mott insulator, governed by a competition between $U$ and $J'$.
However, whether this transition does take place in the bosonic BBH model is not obvious, as strong interactions can potentially lead to a breakdown of the effective theory. 
We now rigorously demonstrate the existence of this exotic phase transition by performing density-matrix renormalization group (DMRG) calculations~\cite{Fishman2022} on a ladder version of the bosonic BBH model.
In Fig.~\ref{fig:BBH}(c), we consider a ladder with $L_y = 2$ sites and up to $L_x = 48$ sites ($N_p = L_x/2$ plaquettes) with filling $\nu = 1/2$, namely two bosons per plaquette. 
The transition is identified by computing the fidelity susceptibility \mbox{$\chi_{_\mathcal F} = -(2/L_x) \lim_{\delta U\to 0} \pa^2 \mathcal F/\pa\delta U^2$} \cite{Zanardi2006,Rigol2013,Sachdev2021}, where $\mathcal F(\delta U) = |\mv{\Psi(U)|\Psi(U+\delta U)}|$.
Our results show the appearance of a clear peak in $\chi_{_\mathcal F}$, located around $U \sim J$ for $J'=0.1J$, which grows with the system size.
For comparison, we solved the effective $p$-band-type model described by Eqs.~\eqref{eq:HamP} and \eqref{eq:projJ1}, and found that the expected transition~\cite{Li2012} indeed occurs in the same range as for the full BBH model [Fig.~\ref{fig:BBH}(c)], thus indicating that the projected theory provides a valid description across the transition.
In addition to the fidelity peak, we also observed the opening of a charge gap \mbox{$\Delta E_c = E(N+1) + E(N-1) - 2E(N)$}, indicating that the system enters an incompressible phase known as chiral Mott insulator~\cite{Li2012, Paramekanti2014}, which is characterized by a finite angular momentum in each plaquette.
This is confirmed by our calculation of the loop current operator $\hat{\mathcal L}_{r} = i \left(\hat b^\dag_{1,r} \hat b^{}_{3,r} + \hat b^\dag_{3,r} \hat b^{}_{4,r} + \hat b^\dag_{4,r} \hat b^{}_{2,r} - \hat b^\dag_{2,r} \hat b^{}_{1,r} - \text{H.c.} \right) \approx \sqrt{2} \hat L_{z,r}$, which remains approximately constant across the transition [Fig.~\ref{fig:BBH}(d))]. 
In contrast with $p$-bands, where angular momentum displays an antiferromagnetic ordering~\cite{Li2012}, the chiral Mott insulator exhibited by the bosonic BBH model exhibits a ferromagnetic ordering.

%\section{Discussion and conclusions}

\emph{Concluding remarks.} Our construction offers the opportunity to observe chiral bosonic phases deep in the strongly-correlated regime, which have remained elusive in higher bands~\cite{Li2012} or in driven systems realizing synthetic flux~\cite{Kolley2015,Greschner2015}. Besides, an interesting perspective concerns the possible interplay between the interacting bosonic phases presented in this work and the unusual topological properties of the underlying BBH band structure~\cite{Benalcazar2017,DiLiberto2020}; see also Refs.~\cite{Neupert2018, Hughes2019, Hatsugai2019, Cuadra2022} on interaction effects in higher-order topological insulators. Our construction can also be applied to other lattice geometries, and extended to higher dimensions, where $\pi$-flux models can exhibit 4-fold ground-state degeneracy~\cite{Benalcazar2017}; see also Refs.~\cite{Polini2005,Mukherjee2018, DiLiberto2019, kremer2020square, zurita2021tunable,Barbiero2022} on various $\pi$-flux models with frustration.

The chiral dynamics studied in our work can also be relevant to nonlinear photonic systems \cite{rechtsman2013photonic,hafezi2013imaging,Ozawa2019,mukherjee2020observation, smirnova2020nonlinear,Carusotto2021}, e.g.~arrays of coupled optical waveguides \cite{Szameit2010}, upon injection of light with the appropriate relative phase pattern among the sites of an elementary plaquette \footnote{Nonlinearities in waveguides are attractive, $g<0$, but we verified that chiral dynamics exists for both positive and negative $U$.}.
This would provide an alternative framework to explore orbital physics \cite{Amo2017,Whittaker2018,Stoferle2021} in the nonlinear regime.

\emph{Acknowledgements.} The authors are pleased to acknowledge discussions with L. Barbiero, I. Carusotto, N. R. Cooper, J. Dalibard, E. Demler, O. K. Diessel, A. Eckardt, D. González-Cuadra, G. Juzeli\=unas, L. Peralta Gavensky, T. Zache and Qi Zhou. 
This work is supported by the ERC Starting Grant TopoCold, the Fonds De La Recherche Scientifique (FRS-FNRS, Belgium), the QuantERA grant MAQS via the Austrian Science Fund FWF No I4391-N and the Rita Levi Montalcini Program through
the fellowship DI\_L\_LEVI22\_01.

\bibliography{biblio}

\clearpage

\newcommand{\beginsupplement}{%
        \setcounter{equation}{0}
        \renewcommand{\theequation}{S\arabic{equation}}%
        \setcounter{figure}{0}
        \renewcommand{\thefigure}{S\arabic{figure}}%
     }
\beginsupplement

\section*{Supplemental Material}

\textbf{Effective theory.}
The single-particle Hamiltonian describing the $\pi$-flux plaquette,
\be
\hat H_0 = -J \left(e^{i\pi}\hat b_1^\dag \hat b_2^{} +  \hat b_2^\dag \hat b_4^{} + \hat b_4^\dag \hat b_3^{} + \hat b_3^\dag \hat b_1^{} + \mathrm{H.c.}\right) \,,
\ee
admits the eigenstates $|d_i\ket = \sum_j \mathcal U_{ij}^{-1} |b_j\ket$, with $i=1,\dots,4$ and 
\be
\mathcal U = 
\renewcommand*{\arraystretch}{1.2}
\begin{pmatrix}
 \frac{1}{2} & -\frac{1}{2} & -\frac{1}{2} & \frac{1}{2} \\
 0 & \frac{1}{\sqrt{2}} & 0 & \frac{1}{\sqrt{2}} \\
 \frac{1}{\sqrt{2}} & 0 & \frac{1}{\sqrt{2}} & 0 \\
 \frac{1}{2} & \frac{1}{2} & -\frac{1}{2} & -\frac{1}{2} \\
\end{pmatrix}\,.
\ee 
The corresponding eigenvalues are $\epsilon_{1,2} = -\sqrt 2 J$ and $\epsilon_{3,4} = \sqrt 2 J$. 
The single-particle gap $\Delta \epsilon = 2\sqrt 2 J$, together with the weak coupling condition $g/J \ll 1$ considered in this work, allows us to project the full interacting Hamiltonian onto the lowest energy single-particle states and to neglect virtual processes involving the high energy modes. 
Specifically, this projection is achieved by decomposing the operators as $\hat b_i^{} = \sum_{ij} \mathcal{U}_{ij} \hat d_j^{}$ and by dropping the contributions from $\hat d_{3,4}$. 
We note that these terms would be relevant to construct an effective theory in second order perturbation theory (here, we limit ourselves to the lowest order contribution). The Hubbard interaction 
\be
\hat H_{\mathrm{int}} = \f{U}{2} \sum_i \hat n_i (\hat n_i - 1),
\ee
thus becomes
\begin{align}
\label{eq:int1}
\hat H_{\textrm{int}}^{\textrm{eff}} \equiv &\, \hat P\hat H_{\textrm{int}} \hat P = \f{3U}{16} \left( \hat d^{\dag}_{1} \hat d^{\dag}_{1} \hat d^{}_{1}\hat d^{}_{1} + \hat d^{\dag}_{2} \hat d^{\dag}_{2} \hat d^{}_{2}\hat d^{}_{2} \right) \\
& + \f U 4  \hat d^{\dag}_{1} \hat d^{\dag}_{2} \hat d^{}_{1}\hat d^{}_{2} 
+ \f{U}{16} \left( \hat d^{\dag}_{1} \hat d^{\dag}_{1} \hat d^{}_{2}\hat d^{}_{2} +\hat d^{\dag}_{2} \hat d^{\dag}_{2} \hat d^{}_{1}\hat d^{}_{1}\right) \nn \,,
\end{align}
where $\hat P$ is the projection operator. Note the orbital-changing terms in the second line of Eq.~\eqref{eq:int1}. 
Let us define the density operator $\hat n = \hat d_1^\dag \hat d_1^{} + \hat d_2^\dag \hat d_2^{}$ and the angular momentum operator $\hat L_z = i (\hat d_1^\dag \hat d_2^{} - \hat d_2^\dag \hat d_1^{})$. 
The interaction Hamiltonian finally reads
\be 
\hat H_{\textrm{int}}^{\textrm{eff}} = \f{3U}{16} \hat n^2  - \f{U}{16} \hat L_z^2 - \f U 8 \hat n\,.
\ee 
As $[\hat n, \hat L_z] = 0$, the eigenstates of $\hat H_{\textrm{int}}^{\textrm{eff}}$ can be written in terms of the eigenstates of $\hat L_z$, which we indicate as $|\pm \ket \equiv \hat d^\dag_\pm |0\ket$, where $\hat d^\dag_\pm = (\hat d^\dag_1 \pm i \hat d^\dag_2)/\sqrt 2$, and satisfying $\hat L_z |\pm\ket = \pm |\pm\ket$.
A generic many-body eigenstate can therefore be written as 
\be
|n_+, n_- \ket = \f{1}{\sqrt{n_+!\, n_-!}}(\hat d_+^\dag)^{n_+} (\hat d_-^\dag)^{n_-} |0\ket\,,
\ee
with $N = n_+ + n_-$. The corresponding eigenvalues read 
\be
E(n_+,n_-) = \f{3U}{16} N^2 - \f{U}{16} ( n_+-n_-)^2 - \f U 8 N\,.
\ee
As only one between $n_-$ and $n_+$ is independent, there are in total $N+1$ states. 
Inspection of this expression shows that there are two degenerate ground states corresponding to $n_+ = N$ or $n_- = N$ with interaction energy $E_{\rm{GS}} = N g/8 - g/8$.
These results correspond to the mean-field results, detailed in the main text, for $N\ra\infty$.
As here we are solving for the exact eigenstates, we also have found a beyond-mean field correction to the energy that does not scale with the number of particles $N$.
Let us consider the ground state with $n_+ = N$ and $n_- = 0$.
In the particle-conserving framework, the lowest-energy excitation corresponds to a single-particle excitation  of the form $n_+' = n_+ - 1 = N - 1$ and $n_-' = n_- +1 = 1$, namely a particle transferred to a state with opposite chirality.
This corresponds to a change in the angular momentum, and the energy for this excitation is 
\be
\Delta E = \f g 4 - \f U 4 > 0\,.
\ee
%%%%%%%%%  Figure  %%%%%%%%%%%%%%%%%%%%%%%%%
\begin{figure}[!t]
\center
\includegraphics[width=\columnwidth]{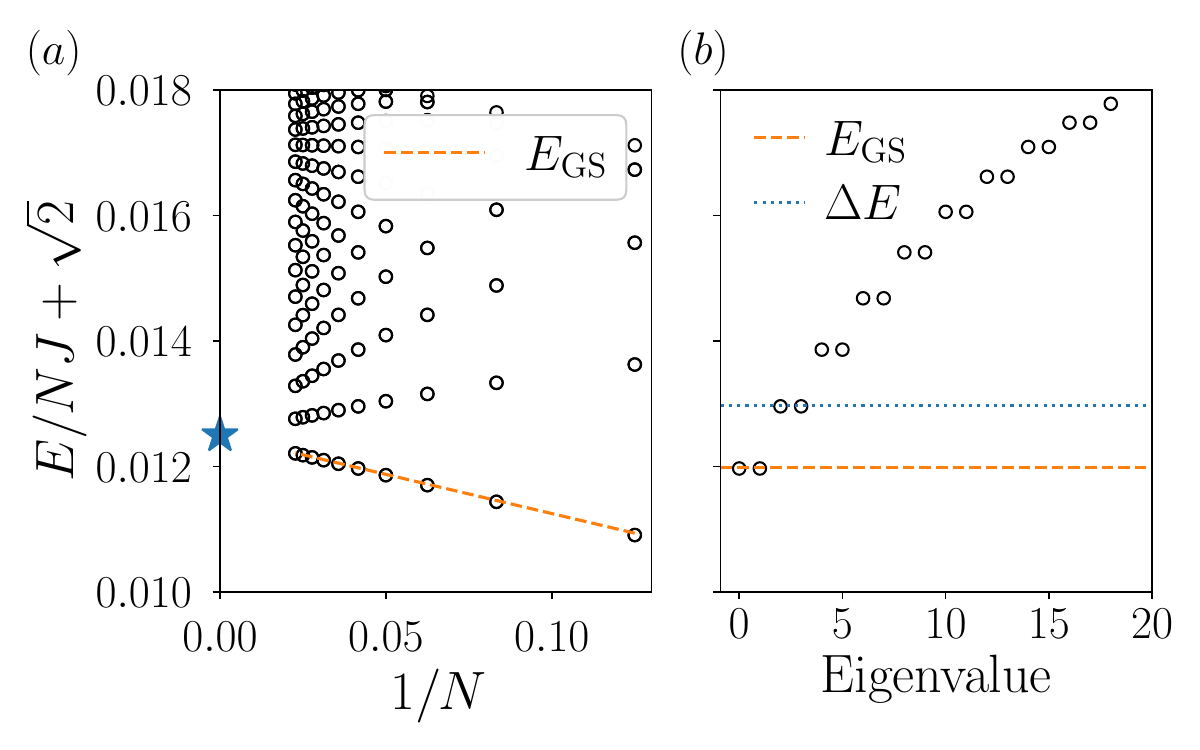}

\caption{Spectrum of the effective model. (a) Energy spectrum obtained with exact diagonalization for $g=UN=0.1J$ as a function of number of particles $N$. The dashed line is the analytical ground state energy $E_{\text{GS}}$ (see text), whereas the star is the mean-field result for $N\rightarrow\infty$. (b) Spectrum for $N=24$ bosons. The dashed and dotted lines indicate the analytical values of the ground state energy $E_\text{GS}=Ng/8-g/8$ and of the lowest energy excitation $\Delta E = g/4 - U/4$.
}
\label{fig:energy_GS}
\end{figure}
This is exactly the result that we obtained within the Bogoliubov theory for the gapped mode.
We therefore have identified that the gapped mode describes a single particle excitation corresponding to flipping the angular momentum of a particle in the condensate.
The mean-field equations of motion can be derived from the Lagrangian \eqref{eq:lagrangian}, which read
\begin{align}
i\pa_t d_1 &= \f{3U}{8} (n_1 + n_2) d_1 - i \f U 8 L_z d_2 - \f{V(t)}{2} d_1  \,,\nn\\
i\pa_t d_2 &= \f{3U}{8} (n_1 + n_2) d_2 + i \f U 8 L_z d_1 + \f{V(t)}{2} d_2\, ,
\label{eq:GPeff}
\end{align}
where we included also the drive $V(t)$. These equations were numerically solved in the presence of the drive to excite the collective mode via a time-periodic modulation.

\begin{figure}[!t]
\center
\includegraphics[width=0.8\columnwidth]{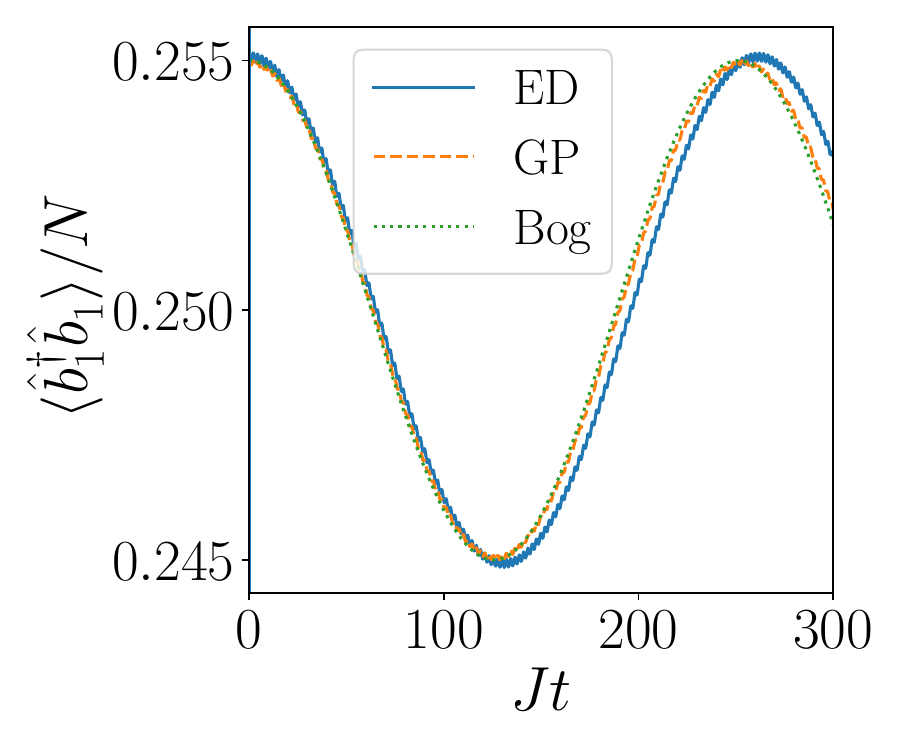}

\caption{Comparison of the impurity dynamics obtained from exact diagonalization, Gross-Pitaevskii and Bogoliubov theory. Parameters as in the main text.
}
\label{fig:comparison_GPEDB}
\end{figure}

\textbf{Bogoliubov theory.} Here we show how Bogoliubov theory can be used to describe the 4-sites GP dynamics for the excited mode.
After replacing $\hat d_1 = \mv{\hat d_1} + \delta \hat d_1$, $\hat d_2 = \mv{\hat d_2} + \delta \hat d_2$ in $\hat H_\text{eff}$ and analogous relations for the Hermitian conjugates, we obtain the single-plaquette Bogoliubov Hamiltonian $\hat H_{\text{Bog}} = \om_0\, \hat \beta^\dag \hat\beta$, where $\hat\beta^\dag = (i\, \delta \hat d_1^\dag + \delta \hat d_2^\dag)/\sqrt 2$. 
In terms of the sites fluctuation operators $\delta \hat b_i^{(\dag)}$, the Bogoliubov Hamiltonian governing the dynamics of the excitations reads
\be
\hat H_{\text{Bog}} = \sum_{i,j} \left( \tilde J_{ij} \delta \hat b_i^\dag \hat \delta b_j^{} + \text{H.c.} \right)\,,
\ee
where the nearest-neighbor couplings are $\tilde J_{21} \!=\! -(\om_0/4) e^{i\pi/4}$, $\tilde J_{13} \!=\! \tilde J_{34} \!=\! \tilde J_{42} \!=\! - \tilde J_{21}$ and the next-nearest-neighbor ones $\tilde J_{14} \!=\! \tilde J_{32} \!=\! (\om_0/4) e^{i\pi/2}$. 
Notice the appearance of couplings between nearest- and next-nearest-neighbor sites that are proportional to $\om_0$. 
We now evolve the mean-field initial impurity state, obtained by adding an onsite potential $-\Delta \hat b^\dag_{1} \hat b^{}_{1}$ as in the main text to find the ground state and then quenching $\Delta \ra 0$, with such free Hamiltonian.
The results of the dynamics under $\hat H_{\text{Bog}}$ is shown in Fig.~\ref{fig:comparison_GPEDB} and agrees with the GP and ED results.

%%%%%%%%%  Figure  %%%%%%%%%%%%%%%%%%%%%%%%%

\textbf{Decay rate of the gapped mode.} Under the mean-field substitution $\hat d_{1,\b r} \ra  \sqrt{\rho/2} + \delta \hat d_{1,\b r}$ and $\hat d_{2,\b r} \ra  i \sqrt{\rho/2} + \delta \hat d_{2,\b r}$, the BBH interacting model can be decomposed into $\hat H = \hat H^{(0)} + \hat H^{(2)} + \hat H^{(3)}$, where each term $\hat H^{(i)}$ is defined by having the corresponding powers of operators $\delta \hat d_{\sigma,\b r}^{(\dag)}$. 
The linear term vanishes by choosing the chemical potential accordingly. 
The quadratic (Bogoliubov approximation) term $\hat H^{(2)}$ reads
\begin{widetext}
\begin{align}
\hat H^{(2)} = & - \f{J'}{2 \sqrt 2} \sum_{\b r, \nu}  \left( \delta \hat d^{\dag}_{1, \b r} \delta \hat d^{}_{1,\b r + \b e_{\nu}} + \textrm{H.c.} \right)  - \f{J'}{2 \sqrt 2} \sum_{\b r, \nu}  \left( \delta \hat d^{\dag}_{2, \b r} \delta \hat d^{}_{2,\b r + \b e_{\nu}} + \textrm{H.c.} \right) \nn \\
& -\sqrt 2 \, J \sum_{\b r}  \left( \delta \hat d^{\dag}_{1, \b r} \delta \hat d^{}_{1,\b r} + \delta \hat d^{\dag}_{2, \b r} \delta \hat d^{}_{2,\b r} \right) - \mu  \sum_{\b r}  \left( \delta \hat d^{\dag}_{1, \b r} \delta \hat d^{}_{1,\b r} + \delta \hat d^{\dag}_{2, \b r} \delta \hat d^{}_{2,\b r} \right) \nn \\
& + \f g 2   \sum_{\b r}  \left( \delta \hat d^{\dag}_{1, \b r} \delta \hat d^{}_{1,\b r} + \delta \hat d^{\dag}_{2, \b r} \delta \hat d^{}_{2,\b r} \right) + \f{g}{16}  \sum_{\b r}  \left( \delta \hat d^{\dag}_{1, \b r} \delta \hat d^{\dag}_{1,\b r} + \delta \hat d^{}_{1, \b r} \delta \hat d^{}_{1,\b r} \right) \nn\\
& - \f{g}{16}  \sum_{\b r}  \left( \delta \hat d^{\dag}_{2, \b r} \delta \hat d^{\dag}_{2,\b r} + \delta \hat d^{}_{2, \b r} \delta \hat d^{}_{2,\b r} \right) + i\f{g}{8}  \sum_{\b r}  \left( \delta \hat d^{\dag}_{1, \b r} \delta \hat d^{\dag}_{2,\b r} - \delta \hat d^{}_{1, \b r} \delta \hat d^{}_{2,\b r} \right) \,.
\end{align}
that can be recast in momentum space after introducing the fluctuation field operator $\delta \hat \Psi_{\b k}^{} \equiv (\delta \hat d_{1,\b k}^{}, \delta \hat d_{2,\b k}^{}, \delta \hat d_{1,-\b k}^{\dag}, \delta \hat d_{2,-\b k}^{\dag})^T$ in order to obtain the Bogoliubov Hamiltonian $\hat H^{(2)} = \f 1 2 \sum_{\b k} \delta \hat \Psi_{\b k}^{\dag} \, H(\b k) \, \delta \hat \Psi_{\b k}^{}$
where 
\be
H(\b k) =
\begin{pmatrix}
\Sigma_{\b k} & \Delta \\
\Delta^\dag & \Sigma_{\b k}
\end{pmatrix}\,, \,\, \textrm{with} \,\,\,
\Delta = \f{g}{8}\,
\begin{pmatrix}
1 & i \\
i  & - 1
\end{pmatrix}\,,
\ee
and $\Sigma_{\b k} \!=\! (\epsilon_{\b k} \!-\! \mu \!+\! g/2)\,  \mathcal I_{2\times2}$, $\epsilon_{\b k} \!=\! -\sqrt{2} J -J'(\cos k_x a  \!+\! \cos k_y a)/\sqrt 2$ and $\mu \!=\! -\sqrt 2 (J\!+\!J') \!+\! g/4$\,. 
Diagonalization yields $\hat H^{(2)} = \sum_{\b k\neq 0, \alpha} \om_{\alpha,\b k}\,  \hat \beta^{\dag}_{\alpha,\b k} \hat \beta^{}_{\alpha,\b k}$, with $\alpha=1,2$, where $\hat \beta^{\dag}_{\alpha,\b k}$ creates a Bogoliubov quasiparticle and the spectrum reads  $\om_{1,\b k} = \sqrt{\xi_{\b k}(\xi_{\b k}+ g )}$ and $\om_{2,\b k} = g/4 +  \xi_{\b k}$
with $\xi_{\b k} \equiv -J'(\cos k_xa + \cos k_ya - 2)/\sqrt 2 $. 

For the decay of the gapped mode, we are interested in the cubic term, $\hat H^{(3)}$, that couple Bogoliubov quasiparticles of different branches thus determining the leading channel for the the decay of the excited mode.
The cubic order reads 
\begin{align}
\f{16}{U}  \sqrt{\f 2 \rho} \hat H^{(3)} & = \,  +3 \,  \delta \hat d^\dag_{1,\b r} \delta \hat d^{}_{1,\b r} \delta \hat d^{}_{1,\b r} -3i  \, \delta \hat d^\dag_{1,\b r} \delta \hat d^{}_{1,\b r} \delta \hat d^{}_{2,\b r} -i\, \delta \hat d^\dag_{1,\b r} \delta \hat d^{}_{2,\b r} \delta \hat d^{}_{1,\b r} + \, \delta \hat d^\dag_{1,\b r} \delta \hat d^{}_{2,\b r} \delta \hat d^{}_{2,\b r}  \nn \\
& \quad -i\, \delta \hat d^\dag_{2,\b r} \delta \hat d^{}_{1,\b r} \delta \hat d^{}_{1,\b r} -\, \delta \hat d^\dag_{2,\b r} \delta \hat d^{}_{1,\b r} \delta \hat d^{}_{2,\b r} + 3\, \delta \hat d^\dag_{2,\b r} \delta \hat d^{}_{2,\b r} \delta \hat d^{}_{1,\b r} -3i\, \delta \hat d^\dag_{2,\b r} \delta \hat d^{}_{2,\b r} \delta \hat d^{}_{2,\b r} \nn \\
& \quad + 3 \,  \delta \hat d^{}_{1,\b r} \delta \hat d^\dag_{1,\b r} \delta \hat d^{}_{1,\b r} +i  \, \delta \hat d^{}_{1,\b r} \delta \hat d^\dag_{1,\b r} \delta \hat d^{}_{2,\b r} -i\, \delta \hat d^{}_{1,\b r} \delta \hat d^\dag_{2,\b r} \delta \hat d^{}_{1,\b r} +3 \, \delta \hat d^{}_{1,\b r} \delta \hat d^\dag_{2,\b r} \delta \hat d^{}_{2,\b r} \nn \\
& \quad -3i\, \delta \hat d^{}_{2,\b r} \delta \hat d^\dag_{1,\b r} \delta \hat d^{}_{1,\b r} +\, \delta \hat d^{}_{2,\b r} \delta \hat d^\dag_{1,\b r} \delta \hat d^{}_{2,\b r} - \, \delta \hat d^{}_{2,\b r} \delta \hat d^\dag_{2,\b r} \delta \hat d^{}_{1,\b r} -3i\, \delta \hat d^{}_{2,\b r} \delta \hat d^\dag_{2,\b r} \delta \hat d^{}_{2,\b r} + \rm{H.c.} \nn\\
& = \sum_{ijk} A_{ijk} \, \delta \hat d^\dag_{i,\b r} \delta \hat d^{}_{j,\b r} \delta \hat d^{}_{k,\b r} + B_{ijk} \, \delta \hat d^{}_{i,\b r} \delta \hat d^\dag_{j,\b r} \delta \hat d^{}_{k,\b r} + C_{ijk} \, \delta \hat d^\dag_{i,\b r} \delta \hat d^\dag_{j,\b r} \delta \hat d^{}_{k,\b r} + D_{ijk} \, \delta \hat d^\dag_{i,\b r} \delta \hat d^{}_{j,\b r} \delta \hat d^\dag_{k,\b r} \nn \,,
\end{align}
where $C_{ijk} = A_{kji}^*$ and $D_{ijk} = B_{kji}^*$.
\end{widetext}
Let us denote the Bogoliubov transformation $\delta \hat d^{}_{i,\b k} = u_{ij,\b k} \hat \beta_{j,\b k} + v^*_{ij,\b k} \hat \beta^\dag_{j,-\b k}$, we therefore find 
\begin{align}
\f{16}{U} \sqrt{\f 2 \rho} \hat H^{(3)} = \f{1}{N_s^{1/2}} \sum_{\b q} \hat \beta^{}_{2,\b 0} \hat \beta^\dag _{1,\b q}\hat \beta^\dag_{1,-\b q} \, F_{\b q} + \dots
\end{align}
where $N_s$ is the number of sites and we retained only the terms determining the decay $|\beta_{2,\b 0}\ket \ra | \beta_{1,\b q} \beta_{1,-\b q} \ket $ and defined $ F_{\b q} = A_{ijk} P_{ijk,\b q} + B_{ijk} Q_{ijk,\b q} + C_{ijk} R_{ijk,\b q} + D_{ijk} S_{ijk,\b q}$, where
\begin{align}
P_{ijk,\b q} &= u^*_{i1,\b q} u^{}_{j2,\b 0} v^*_{k1,\b q} + u^*_{i1,\b q} v^*_{j1,-\b q} u^{}_{k2,\b 0} \nn \\
Q_{ijk,\b q} &= u^{}_{i2,\b 0} u^*_{j1,\b q} v^*_{k1,\b q} + v^*_{i1,\b q} u^*_{j1,\b q} u^{}_{k2,\b 0} \nn \\
R_{ijk,\b q} &= u^*_{i1,\b q} u^*_{j1,-\b q} u^{}_{k2,\b 0} \nn \\
S_{ijk,\b q} &= u^*_{i1,\b q} u^{}_{j2,\b 0} u^*_{k1,-\b q} \,.
\end{align}
For $g \ll J'$, when the gap is much smaller than the bandwidth, we are in the small wavelength limit and the Bogoliubov Hamiltonian has the form
\be
H^{(2)}_{\tilde{\b q}} = 
\left(
\begin{array}{cccc}
 \tilde q^2+\omega_0  & 0 & \om_0/2 & i \om_0/2 \\
 0 & \tilde q^2+\omega_0  & i\om_0/2 & -\om_0/2 \\
 \om_0/2 & -i\om_0/2 & \tilde q^2+\omega_0  & 0 \\
 -i\om_0/2 & -\om_0/2 & 0 & \tilde q^2+\omega_0  \\
\end{array}
\right)\,,
\ee
where we defined $\tilde q^2 = q^2 (J'/2\sqrt 2)$ and $q^2 = q_x^2 + q_y^2$.
The spectrum of $\Sigma_z H^{(2)}_{\tilde{\b q}}$ reads $\epsilon_{1,\tilde{\b q}} = \sqrt{\tilde q^2 (\tilde q^2 + 2\om_0)}$ and $\epsilon_{2,\tilde{\b q}} = \omega_0 + \tilde q^2$, whereas the Bogoliubov modes are 
\be
u_{ij, \tilde{\b q}} = 
\begin{pmatrix}
-i u_{\tilde q} & \f{i}{\sqrt 2} \\
u_{\tilde q} & \f{1}{\sqrt{2}}
\end{pmatrix}\,, \qquad
v_{ij, \tilde{\b q}} = 
\begin{pmatrix}
i v_{\tilde q} & 0 \\
v_{\tilde q} & 0
\end{pmatrix}\,,
\ee
with $u^2_{\tilde q} = \f{\xi^2_{\tilde q}}{2(\xi^2_{\tilde q}-1)}$ and $v^2_{\tilde q} = \f{1}{2(\xi^2_{\tilde q}-1)}$, and we defined $\xi_{\tilde q} = (\epsilon_{1,\tilde q}+\epsilon_{2,\tilde q})/\omega_0$.
By direct calculation, we find that the only nonvanishing contribution to $F_q$ is \mbox{$F_{\b q} = \sum_{ijk} C_{ijk} R_{ijk,\b q} = -i\sqrt 2 u_{q}^2$}.

The decay rate of the gapped mode can now be calculated by [S1] 
\be 
\Gamma = \pi \sum_{\b q} \delta (\om_0 - 2 \epsilon_{1,\b q})\, |\bra \beta_{1,\b q} \beta_{1,-\b q} | \hat H^{(3)} |  \beta_{2, \b 0} \ket|^2\,,
\ee
and the resonance condition is satisfied when \mbox{$\tilde q^2 = (\sqrt 2 -1) \om_0$} yielding $\xi_{\tilde q}= 1 + \sqrt 2$. 
After transforming $N_s^{-1/2}\sum_{\b q} \ra 2\pi\int \rm d q \, q$, we obtain
\begin{align}
\Gamma  &= \f{\sqrt 2\pi^2 g U}{128 J'} \int \rm d \tilde q \, \tilde q \, |F_{\tilde q}|^2 \delta (\om - 2 \epsilon_{1,\tilde q}) \nn \\
&\approx \f{0.73\pi^2 g U }{ 512 J'}\,.
\end{align}

\textbf{Details on DMRG simulations.} We have used the ITensor library to perform the DMRG calculations.
All the simulations have been obtained by setting the truncation error threshold to $10^{-10}$. 
The bond dimension threshold has been taken up to $\chi = 2000$ for some simulations that required a more accurate convergence and we have used up to 60 sweeps for the ground state search.
In order to select a specific time-reversal broken ground state from the degenerate manifold, we have added a small uniform flux via a complex tunneling phase.

The fidelity susceptibility 
\begin{equation}\nonumber
\chi_{_\mathcal F} = -(2/L_x) \lim_{\delta U\to 0} \pa^2 \mathcal F/\pa\delta U^2\,,
\end{equation}
shown in Fig.~\ref{fig:BBH}(c) of the main text has been computed with a parabolic fit in $\delta U$ of the fidelity $\mathcal F(\delta U) \equiv |\mv{\Psi(U)|\Psi(U+\delta U)}|$ for 10 values of $\delta U < 10^{-3}J$. 

The charge gap shown in Fig.~\ref{fig:BBH}(d) \mbox{$\Delta E_c = E(N+1) + E(N-1) - 2E(N)$}, where $N$ is the number of bosons, has been extrapolated from a linear fit in $1/L_x$, for $L_x = 32, 36, 40, 44, 48$. 

Furthermore, we have checked that across the Mott transition, the spectrum remains doubly degenerate by computing the energy of few low-lying states.

\begin{center}
\noindent\rule{0.1\textwidth}{0.6pt}
\end{center}

\vspace{1cm}
[S1] L. Pitaevskii and S. Stringari, \emph{Bose–Einstein Condensation and Superfluidity}, Oxford University Press (2016)

\end{document}